\documentclass[12pt]{elsarticle}
\usepackage{graphicx}
\usepackage{subfigure}
\usepackage{amsbsy}
\usepackage{times}
\usepackage{gensymb}

\usepackage{caption}
\usepackage{color}
\graphicspath{{fig/}}

\usepackage{graphicx,graphics}
\usepackage{epsfig}
\usepackage{amsfonts,amssymb,amsmath,amsthm}
\usepackage{commath}
\usepackage{amsopn}

\newtheorem*{remark}{Remark}

\usepackage{hyperref}
\usepackage{cleveref}
\usepackage{tikz}
\usepackage{pgfplots}

\pgfplotsset{compat=newest}
\usepackage{filecontents}
\usepackage{lineno}
\usepackage{xspace}
\usepackage{array}
\usepackage[font = normal, labelfont = bf]{caption}
\usepackage[font = normal,figurename = Fig.]{caption}
\usepackage{caption}
\usepackage{multirow}
\usepackage{enumerate}
\usepackage{mathrsfs}

\usepackage{subfigure}
\usepackage{rotating}
\usepackage{empheq}
\usepackage{listings}
\usepackage{dcolumn}
\usepackage{color}
\usepackage{natbib}
\usepackage{booktabs}
\usepackage{diagbox}

\begin{document}
\newcommand{\ii}{\int\limits_{-\infty}^{\infty}}
\def\al{\alpha}
\def\om{\omega}
\def\ba{\begin{array}}
\def\ea{\end{array}}
\def\vcap{\overline{v}}
\def\eqref#1{(\ref{#1})}
\newcommand{\bT}{{\mbox{\boldmath ${\rm T}$}}}
\newcommand{\bY}{{\mbox{\boldmath ${\rm Y}$}}}
\newcommand{\bS}{{\mbox{\boldmath ${\rm S}$}}}
\newcommand{\btau}{{\mbox{\boldmath $\tau$}}}
\begin{frontmatter}
\title{Polytopal composite finite elements for modeling concrete fracture based on nonlocal damage models}

\author[ath1]{Hai D. Huynh}
\author[india]{S. Natarajan}
\author[ath2]{H. Nguyen-Xuan\corref{cor1}}
\ead{ngx.hung@hutech.edu.vn}
\author[ath1]{Xiaoying Zhuang\corref{cor1}}
\ead{zhuang@hot.uni-hannover.de}

\address[ath1]{Chair of Computational Science and Simulation Technology, Department of Mathematics and Physics, Leibniz University Hannover, Germany}
\address[india]{Integrated Modelling and Simulation Lab, Department of Mechanical Engineering, Indian Institute of Technology Madras, Chennai-600036, India}
\address[ath2]{CIRTech Institute, Chi Minh City University of Technology (HUTECH), Ho Chi Minh City, Vietnam}

\cortext[cor1]{Corresponding author.}

\begin{abstract}
The paper presents an assumed strain formulation over polygonal meshes to accurately evaluate the strain fields in nonlocal damage models. An assume strained technique based on the Hu-Washizu variational principle is employed to generate a new strain approximation instead of direct derivation from the basis functions and the displacement fields. The underlying idea embedded in arbitrary finite polygons is named as \textit{Polytopal composite finite elements} (PCFEM). The PCFEM is accordingly applied within the framework of the nonlocal model of continuum damage mechanics to enhance the description of damage behaviours in which highly localized deformations must be captured accurately. This application is helpful to reduce the mesh-sensitivity and elaborate the process-zone of damage models. Several numerical examples are designed for various cases of fracture to discuss and validate the computational capability of the present method through comparison with published numerical results and experimental data from the literature.
\end{abstract}
%

\end{frontmatter}
\section{Introduction}
The recent development of polygonal finite elements has provided an efficient tool for mesh generation and accurate solutions in the engineering simulation. Their applications have been succeeded in various mechanics problems such as analysis of granular materials \cite{Falco_Cola_Petrinic_2017}, incompressible fluid flow \cite{Talisch_Pereira_2014}, polycrystalline materials \cite{Ghosh_Moorthy_1995}, contact models \cite{Khoei_Yasbolaghi_2015}, to name a few. As an applicable technique for meshing complicated geometries, several numerical methods, namely the Virtual element methods (VEM) \cite{Chi_Veiga_2019,Chi_Veiga_2017,Artioli_Veiga_2017}, the Scaled boundary finite element methods (SBFEM) \cite{Ooi_Song_2017,Pramoda_Ooi_2018}, the smoothed finite element method (SFEM) \cite{Natarajan_Bordas_Ooi_2015} have been developed over such polygonal meshes to deal with challenging issues in solid mechanics. As for the scope of the finite element method (FEM), the performance of polygonal elements with the use of rational basis functions including Wachspress, Laplace, Mean Value Coordinates, however, does not fulfill patch tests. It is because the numerical integral over a polygonal domain is inconsistent with the quadrature rule of polynomials. In particular, taking a domain integral is normally carried out through its division into sub-triangles to apply the standard quadrature rule while the shape functions for interpolations on the element are non-polynomial. One of the significant solutions is the use of gradient correction to produce polynomial consistency for these types of basis functions, reported in Refs. \cite{Talischi_Pereira_Menezes_2015,Chi_Talischi_Pamies_2016}. Alternatively, the linear smoothing technique was presented in \cite{Francis_Bernardin_2017} to produce a new strain field based on a linear smoothing function. The two aforementioned approaches, however, require expensive computational cost due to their complicated algorithms. Recently, piece-wise linear basis functions which are constructed by linear shape functions on triangular elements are successfully developed for polygonal elements \cite{HungNguyen_2017}. Although, the algorithm is mathematically simple, their derivatives are constant on each sub-cell. This leads to the lower accuracy of computed strain field. An assumed strain technique over polygonal finite elements with piece-wise linear interpolations, coined as \textit{Polytopal composite finite elements} (PCFEM) was proposed by Hung \cite{Hung_2019} to overcome the constant strains of sub-cells and to create an enhanced strain field in which the orthogonality condition is satisfied. Futher, the elements pass the patch tests of incompressible problems in two and three dimensions to machine precision. The underlying idea is to produce an assumed strain field derived through a polynomial projection of the compatible strains over the sub-cells sub-divided from a polygonal domain.\\

As a potential numerical method, the exploitation of the PCFEM is still encouraging for solving mechanics problems. The choice of a suitable technique to numerically simulate the damage models has been seen as an open topic for many scientific communities. The damage model is classified as a type of smeared crack approach, which is controlled by a scalar parameter to evaluate the level of damage. In particular, the mechanism occurs in microstructures when a zone of material is degraded. Numerically, there are several common approaches to describe such a problem. The early work on local damage model is troubled with pathological mesh-dependence in the numerical results, or in other words, with vanishing localization of the damage zone. \textcolor{blue}{To alleviate these obstacles, the concept of nonlocal damage models was proposed by Cabot and Bazant \cite{Cabot_Bazant_1987} as an efficient technique for regularizing strain localization. The idea was based on the nonlocal continuum theories introduced by Aifantis \cite{Aifantis_1984}, Erigen \cite{Eringen_1983} who developed it for elastic models. Thereafter, its applications were extended to problems of smeared crack models \cite{Bazant_Lin_1988}, plasticity and damage \cite{Bazant_Jirasek_2002}, wave propagations in nanostructures \cite{Wang_2005}. Physically, the nonlocal approaches are to average, in other words, to homogenize microstructural effects of a material such as heterogeneity or defect on macrostructures}. Herein, a material parameter of length scales represented for the size effects of materials is introduced into the constitutive relations. There are two common types of the nonlocal model for damage mechanics, namely, the non-local integral model \cite{Lorentz_2017,Giry_Dufour_Mazars_2011}, and the gradient-enhanced model \cite{Thai_Rabczuk_Bazilevs_2016,Peerlings_Borst_Brekelmans_1998}. The gradient model is considered as a inherent version of the nonlocal integral type by expansion of the Taylor series \cite{Velde_Kowalsky_Zumendorf_2009,Simone_Askes_Sluys_2004}. Although, the gradient counterpart is more efficient in terms of the computational cost, the main drawback is the treatment of boundary conditions which do not have a physical meaning. \textcolor{blue}{As a excellent solution to damage analysis, much attention have been paid to developing both approaches for fracture mechanics. The most popular object is the use of isotropic continuum damage models \cite{Peerlings_Borst_Brekelmans_1998,Lorentz_2017,Bobinski_Tejchman_2005}to simulate failure in brittle materials, e.g., concrete, rock. These models were accordingly enriched to anisotropic damage in which the damage variable is considered by a tensorial representation \cite{Desmorat_2007,Jin_2018}. Concerning more complicated damage behaviors, creep fracture in polycrystalline materials \cite{Jackiewicz_2007} and ice-sheets \cite{Duddu_Waisman_2013}, and ductile fracture in hyperelastic materials \cite{ Mediavilla_2006} were successfully developed within the nonlocal framework.}\\

In this work, \textcolor{blue}{the theory of quasi-brittle fracture is employed to develop nonlocal continuum damage models of concrete because their mechanical behaviors are relatively identical.} The compatible strain field in the weak form is replaced with an assumed strain field from the PCFEM. The integral-type nonlocal model is then implemented based on the concept of the PCFEM. The nonlocal equivalent strain is defined from the distribution of neighbor or "local" variables inside a characteristic size in which the local equivalent strain is obtained from the assumed strain field. Due to the presence of the damage variable, the constitutive equation becomes nonlinear. Further, the need of local mesh refinements around the damage zones is done by polytree adaptive computations, reported in \cite{Hung_2016}. The cumbersome problem of hanging nodes resulting from the generation of local refinements are handled within the framework of the PCFEM. The applications of PCFEM into nonlocal damage models are encouraging due to the above advantages, in which the desirable accuracy of local variables is achieved by the developed numerical method, and mechanical effects of strongly localized deformations are elaborately handled by the nonlocal formulations. \\

The paper is outlined as follows: the implementation of the assumed strain technique into polygonal finite elements is reported in Section \ref{sec: Polytopal composite finite elements}. \textcolor{blue}{The ability of PCFEM to handle the less accuracy of computations with hanging nodes is discussed in this section.} In Section \ref{sec damage_formulations}, the nonlocal damage formulations defined over PCFEM are presented, followed by mathematical derivation of the nonlocal tangent stiffness matrix. Several numerical examples are designed to validate the reliability of the present method in various damage problems including pure-tension mode and mixed-mode fractures, shown in Section \ref{sec: Numerical examp}.

\section{Polytopal composite finite elements}
\label{sec: Polytopal composite finite elements}
\subsection{Assumed strain on polygonal finite elements}
\label{subsec: assumed strain}

\begin{figure}
	\centering
	\includegraphics[scale=1]{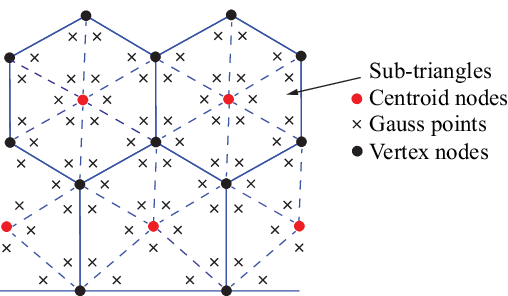}
    \caption{Description of Gauss distributions, centroid nodes in an extraction of a polygonal mesh.}
    \label{fig.Gauss_points}
\end{figure}

With the straightforward procedure to build interpolations and to perform numerical integration over a polygonal domain $\Omega ^e$, the piece-wise linear basis functions are chosen. These shape functions are shown to be an efficient tool to conform to approximations on both convex and non-convex elements and to save computational cost~\cite{HungNguyen_2017,Hung_2019}. Basically, the poly-piece-wise basis functions are constructed through sub-triangles created by connecting the centroid with two adjacent vertices of an $n$-gon, as described in Fig. \ref{fig.Gauss_points}.\\
\\
Their values at these points are defined as
\begin{equation}
        {\phi _i}\left( {\bf{x}} \right) = \left\{ {\begin{array}{*{20}{c}}
	{{\delta _{ij}}{\quad \rm{ if \,}}{\bf{x}}{\rm{ \, = \, }}{{\bf{x}}_j}}\\
	{1/n{\quad \rm{ if \,}}{\bf{x}}{\rm{ \, = \, }}{{\bf{x}}_C}}
	\end{array}} \right.
\end{equation}
The piece-wise basis functions are given by the triangular linear shape functions over the sub-triangles. The global formulation and the first derivatives are expressed as
\begin{equation}
         {\phi _i}\left( {\bf{x}} \right) = \sum\limits_{j = 1}^3 {\phi _j^{T3}\left( {\bf{x}} \right){\phi _i}\left( {{{\bf{x}}_j}} \right)} ,{\rm{ for \,}}{\bf{x}} \in {\Omega ^{T3}}
         \label{Eq.shape_func_PW}
\end{equation}
\begin{equation}
          \nabla {\phi _i}\left( {\bf{x}} \right) = \sum\limits_{j = 1}^3 {\nabla \phi _j^{T3}\left( {\bf{x}} \right){\phi _i}\left( {{{\bf{x}}_j}} \right)} ,{\rm{ for \,}}{\bf{x}} \in {\Omega ^{T3}}
          \label{Eq.dertivative_PW}
\end{equation}
An assumed strain technique is then introduced to generate a new strain field. The assumed strain field $\boldsymbol{\tilde \varepsilon }$ is obtained by projecting the space of compatible strains, ${{\cal S}_c}$ onto a space of polynomial strains, $\tilde {\cal S}$. \textcolor{blue}{The definition of assumed strain fields is expressed through a set of polynomial least-square}. In detail, for two dimensional solids, the assumed strains at a point ${\mathbf{x}} = \left( {x,y} \right)$ in ${\Omega ^h}$, denoted by ${{\boldsymbol{\tilde \varepsilon }}^T}\left( {\bf{x}} \right) =\left[ {{{\tilde \varepsilon }_{xx}}{\rm{ \quad }}{{\tilde \varepsilon }_{yy}}{\rm{\quad  }}{{\tilde \varepsilon }_{xy}}} \right]$are formed as
\begin{equation}
          {{\boldsymbol{\tilde \varepsilon }}^T}\left( {\bf{x}} \right) = {{\boldsymbol{\beta}}_0} + {{\boldsymbol{\beta }}_1}x + {{\boldsymbol{\beta }}_2}y + ... = {\bf{S}}\left( {\bf{x}} \right){\boldsymbol{\beta }}
          \label{Eq.poly_strain}
\end{equation}
in which $\boldsymbol{\beta}$ is unknown strain fields expressed in the following matrix
\begin{equation}
          {\boldsymbol{\beta }} = \left[ {\begin{array}{*{20}{c}}
	{{\beta _{00}}}&{{\beta _{01}}}&{{\beta _{02}}}\\
	{{\beta _{10}}}&{{\beta _{11}}}&{{\beta _{12}}}\\
	{{\beta _{20}}}&{{\beta _{21}}}&{{\beta _{22}}}\\
	{...}&{...}&{...}
	\end{array}} \right]
          \label{Eq.matrix_strain}
\end{equation}
while ${\bf{S}}\left( {\bf{x}} \right)$ is a vector of the Pascal`s triangle polynomials expressed in the following form:
\begin{equation}
          {\bf{S}}\left( {\bf{x}} \right) =\left\{1 \quad x \quad y \quad \cdots  \right\}
          \label{Eq.Pascal}
\end{equation}
It is noted that an extension to high-order approximation for the strain fields are feasible
by considering higher order terms in Eq. \ref{Eq.poly_strain}
However, in this present study, we employ linear polynomials that satisfy the minimum lower-order approximation of the displacement fields. 
To seek solutions of the assumed strain field,  the error between the $\boldsymbol{\tilde \varepsilon }$ and $\boldsymbol{\varepsilon }$ over each sub-domain is minimized. Note that the space $\tilde {\cal S}$ is mapped from ${{\cal S}_c}$ through a polynomial projection ${\pi _k}$, where $k$ is degree of the polynomial. Thus, the definition of  relation ${\boldsymbol{\tilde \varepsilon }} = {\pi _k}{\boldsymbol{\varepsilon }} \in \tilde {\cal S}$ can be described as:
\begin{equation}
          {\pi _k}{\boldsymbol{\varepsilon }} = \frac{1}{2}\underbrace {{\mathop{\rm argmin}\nolimits} }_{\forall {\boldsymbol{\tilde \varepsilon }} \in \tilde {\cal S}}{\cal E}
          \label{Eq.map_strain}
\end{equation}
where ${\cal E}$ is a functional referring to the error between the assumed and the compatible strain field over a polygonal domain $\Omega_e$, is expressed as:
\begin{equation}
          \textcolor{blue}{
          	{\cal E} = \left\| {{\boldsymbol{\tilde \varepsilon }} - {\boldsymbol{\varepsilon }}} \right\|_{{{\cal L}^2}\left( {{\Omega _e}} \right)}^2 = \int_{{\Omega _e}} {{{\left( {{\boldsymbol{\tilde \varepsilon }} - {\boldsymbol{\varepsilon }}} \right)}^T}} \left( {{\boldsymbol{\tilde \varepsilon }} - {\boldsymbol{\varepsilon }}} \right)d\Omega  = \sum\limits_{i = 1}^{n_T} {\int_{\Omega _i^{T3}} {{{\left( {{\boldsymbol{\tilde \varepsilon }} - {{\boldsymbol{\varepsilon }}^i}} \right)}^T}} \left( {{\boldsymbol{\tilde \varepsilon }} - {{\boldsymbol{\varepsilon }}^i}} \right)d\Omega }
          }
          \label{Eq.error_strain}
\end{equation}
By substituting Eq. \ref{Eq.error_strain} into Eq. \ref{Eq.map_strain}, we have
\begin{equation}
          \frac{{\partial {\cal E}}}{{\partial {\boldsymbol{\beta }}}} = {\bf{0}}
          \label{Eq.deri_strain}
\end{equation}
Mathematically, Eq. \ref{Eq.deri_strain} can be rewritten under the form of an equation system with respect to the unknowns $\boldsymbol{\beta }$
\begin{equation}
          {\bf{M}\boldsymbol{\beta }} = {\bf{Q}}
          \label{Eq.beta}
\end{equation}
The expression for components $\mathbf{M}$ and $\mathbf{Q}$ in Eq. \ref{Eq.beta} is given by:
\begin{equation}
          \textcolor{blue}{
          {\bf{M}} = \sum\limits_{i = 1}^{n_T} {\int_{\Omega _i^{T3}} {{{\bf{S}}^T}\left( {\bf{x}} \right)} } {\bf{S}}\left( {\bf{x}} \right)d\Omega  = \sum\limits_{i = 1}^{n_T} {\sum\limits_{j = 1}^{{n_{GP}}} {\overbrace {\left[ {\begin{array}{*{15}{c}}
				1&x&y\\
				x&{{x^2}}&{xy}\\
				y&{xy}&{{y^2}}\\
				{...}&{...}&{...}
				\end{array}} \right.}^{k = 1}{\rm{   }}\left. {\begin{array}{*{5}{c}}
			{...}\\
			{...}\\
			{...}\\
			{...}
			\end{array}} \right]} } {w_j}\left\| {{\bf{J}}_j^i} \right\|
		}
		\label{Eq.M}
\end{equation}

\begin{equation}
       \textcolor{blue}{
          {\bf{Q}} = \sum\limits_{i = 1}^{n_T} {\int_{\Omega _i^{T3}} {{{\left( {{{\boldsymbol{\varepsilon }}^i}\left( {\bf{x}} \right){\bf{S}}\left( {\bf{x}} \right)} \right)}^T}d\Omega } }  = \sum\limits_{i = 1}^{n_T} {\sum\limits_{j = 1}^{{n_{GP}}} {\left[ {\begin{array}{*{15}{c}}
			{\varepsilon _{xx}^i}&{\varepsilon _{yy}^i}&{\varepsilon _{xy}^i}&{...}\\
			{{x_j}\varepsilon _{xx}^i}&{{x_j}\varepsilon _{yy}^i}&{{x_j}\varepsilon _{xy}^i}&{...}\\
			{{y_j}\varepsilon _{xx}^i}&{{y_j}\varepsilon _{yy}^i}&{{y_j}\varepsilon _{xy}^i}&{...}\\
			{...}&{...}&{...}&{...}
			\end{array}} \right]} } {w_j}\left\| {{\bf{J}}_j^i} \right\|
		}
		\label{Eq.Q}
\end{equation}
where $n_T$ and $n_{GP}$ are the number of sub-cells and Gaussian points per a sub-cell, respectively.

The coefficients $\boldsymbol{\beta }$ in Eq. \ref{Eq.beta} is solved by: ${\boldsymbol{\beta }} = {{\bf{M}}^{ - 1}}{\bf{Q}}$, and then by substituting into Eq. \ref{Eq.poly_strain}, the assumed strain field is computed as:
\begin{equation}
          \begin{array}{l}
              {{{\boldsymbol{\tilde \varepsilon }}}^T}\left( {\bf{x}} \right) = {\bf{S}}\left( {\bf{x}} \right){{\bf{M}}^{ - 1}}{\bf{Q}}\\
          \textcolor{blue}{
              {\rm{         }\qquad\;\ } = {\bf{S}}\left( {\bf{x}} \right){{\bf{M}}^{ - 1}}\left( {\sum\limits_{i = 1}^{n_T} {\sum\limits_{j = 1}^{{n_{GP}}} {{{\bf{S}}^T}\left( {{{\bf{x}}_j}} \right){{\left( {{\bf{B}}_e^i\left( {{{\bf{x}}_j}} \right){{\bf{d}}_e}} \right)}^T}{w_j}\left\| {{\bf{J}}_j^i} \right\|} } } \right)
          }\\
          \textcolor{blue}{
              {\rm{         }\qquad\;\ } = {\bf{d}}_e^T\left( {\sum\limits_{i = 1}^n {\sum\limits_{j = 1}^{{n_{GP}}} {\left( {{\bf{S}}\left( {\bf{x}} \right){{\bf{M}}^{ - 1}}{{\bf{S}}^T}\left( {{{\bf{x}}_j}} \right)} \right){{\left( {{\bf{B}}_e^i\left( {{{\bf{x}}_j}} \right)} \right)}^T}{w_j}\left\| {{\bf{J}}_j^i} \right\|} } } \right)
          }\\
{\rm{         }\qquad\;\ } = {\bf{d}}_e^T{{{\bf{\tilde B}}}_e}\left( {\bf{x}} \right)
           \end{array}
\label{Eq.assumed_strain}
\end{equation}
where, ${\bf{B}}_e^i$ is the differential operator obtained from the strain-displacement relation in solid mechanics, and ${{\bf{d}}_e}$ is the vector of nodal displacements of the elemental domain. The corresponding differential tensor ${{\bf{\tilde B}}_e}$ for the assumed strain method is then given by:
\begin{equation}
        \textcolor{blue}{
          {{\bf{\tilde B}}_e}\left( {\bf{x}} \right) = \sum\limits_{i = 1}^{n_T} {\sum\limits_{j = 1}^{{n_{GP}}} {\left( {{\bf{S}}\left( {\bf{x}} \right){{\bf{M}}^{ - 1}}{{\bf{S}}^T}\left( {{{\bf{x}}_j}} \right)} \right){{\left( {{\bf{B}}_e^i\left( {{{\bf{x}}_j}} \right)} \right)}^T}{w_j}\left\| {{\bf{J}}_j^i} \right\|} }
        }
          \label{Eq.assumed_B}
\end{equation}

Upon computing the new differential operator ${{\bf{\tilde B}}_e}$, the numerical evaluation of the stiffness matrix is computed similar to that of the conventional polygonal finite element method (PFEM). \\

\begin{remark}
The accuracy and the convergence properties of the PCFEM have been studied in detail for two and three dimensional problems in \cite{Hung_2019}. In addition, the orthogonality condition arising from the Hu-Washizu principle is satisfied with assumed strain field, expressed in Eq. \ref{Eq.assumed_strain}. 
\end{remark}

Before proceeding with the framework for damage models, the accuracy and the convergence properties of the PCFEM are studied for a singular perturbation method when dealing with elements having hanging nodes as a results of adaptive refinement. 

\subsection{Convergence study}
\label{subsec convergence study}
\begin{figure}[!httb]
	\centering
	\includegraphics[scale=1]{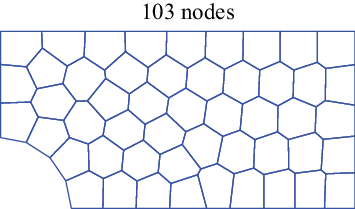}
	\hspace*{1.5em}
	\includegraphics[scale=1]{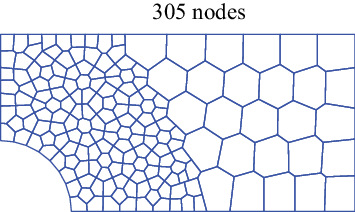}\\
	\vspace*{1em}
	\includegraphics[scale=1]{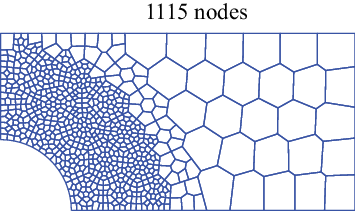}
	\caption{Polygonal meshes in a quarter of the plate.}
	\label{Fig.plate_hole_mesh}
\end{figure}

\begin{figure}[!httb]
	\centering
	\subfigure[]{
		\includegraphics[scale=1]{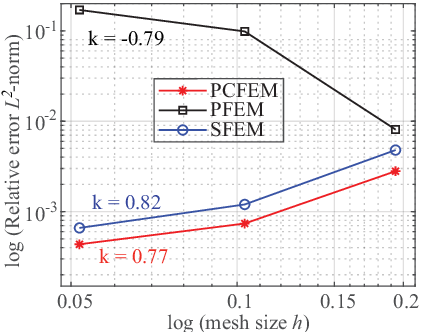}
		\label{Fig.plate_L2}
	} \hspace*{1em}
	\subfigure[]{
		\includegraphics[scale=1]{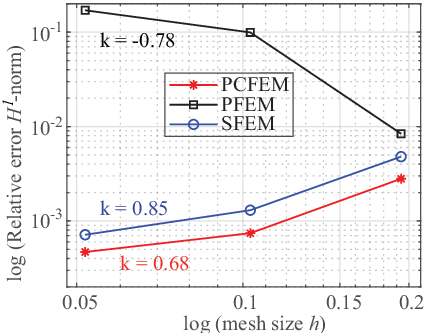}
		\label{Fig.plate_H1}
	}
	\caption{\textcolor{blue}{Relative error of $L^2$-norm (a), and $H^1$-norm (b), obtained from polygonal meshes above.}}
	\label{Fig.convergence_mesh}
\end{figure}

A well-known example regarding a singular perturbation problem is an infinite elastic plate with a center hole, given in Ref. \cite{Meyer_Sayir_1995} is taken into the consideration. This benchmark focuses on the phenomenon of stress concentrations which requires extremely dense meshes to accurately capture the change in the gradient of the displacement field. Fig. \ref{Fig.plate_hole} shows the geometry and the boundary conditions for the problem. \textcolor{blue}{The size of the plate is set up to be sufficiently and much higher than that of the hole. This condition is to ensure the domination of the uniform tension in the far field}. Owing to symmetry, only one quarter of the domain is considered for the analysis. The closed-form solutions defined in the polar coordinates are given by:
\begin{figure}[!httb]
	\centering
	\subfigure[]{
		\includegraphics[scale=1]{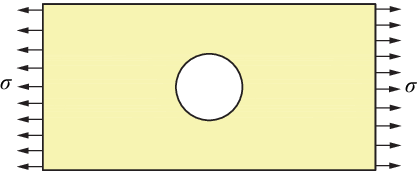}
		\label{Fig.plate_full}
	} \hspace*{1em}
	\subfigure[]{
		\includegraphics[scale=1]{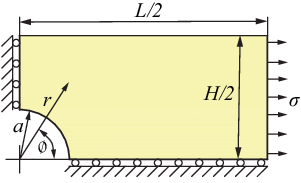}
		\label{Fig.plate_quarter}
	}
	\caption{(a) Full geometry, (b) quarter geometry and boundary conditions of the plate with a hole.}
	\label{Fig.plate_hole}
\end{figure}
\begin{subequations}
	\begin{equation}
    	{\sigma _{rr}} = \frac{\sigma }{2}\left[ {1 - \frac{{{a^2}}}{{{r^2}}} + \left( {1 - \frac{{4{a^2}}}{{{r^2}}} + \frac{{3{a^4}}}{{{r^4}}}} \right)\cos 2\phi } \right]
	\end{equation}
	\begin{equation}
        {\sigma _{\phi \phi }} = \frac{\sigma }{2}\left[ {1 + \frac{{{a^2}}}{{{r^2}}} - \left( {1 + \frac{{3{a^4}}}{{{r^4}}}} \right)\cos 2\phi } \right]
	\end{equation}
	\begin{equation}
	    {\sigma _{r\phi }} =  - \frac{\sigma }{2}\left( {1 + \frac{{2{a^2}}}{{{r^2}}} - \frac{{3{a^4}}}{{{r^4}}}} \right)\sin 2\phi
	\end{equation}
	\label{Eq.Ana_hole}
\end{subequations}
The geometrical parameters for the quarter model are chosen as: $H/2 \times L/2 = 1 \times 2\,m$ with a circular hole $a =$ 0.4 m, and the tensile load is given by $\sigma=$10 kN/m. The material property is chosen as $E=$ 210 GPa, $\nu=$ 0.33. Three different densities of polygonal meshes included with poly-tree mesh refinements as depicted in Fig. \ref{Fig.plate_hole_mesh} are employed to analyze this model.\\
A convergence study is done to assess the accuracy of the solutions obtained from the present method (PCFEM) in dealing with the singular problem induced by hanging nodes in adaptive mesh refinement. Two other numerical methods, namely the polygonal finite element method with Wachspress shape functions (PFEM), and the cell-based smoothed finite element method (SFEM) are used for comparison. As observed, the PCFEM yields highly accurate solutions in terms of the relative error of $L^2$-norm and $H^1$-norm plotted in \textit{log-log} scale as shown in Fig. \ref{Fig.convergence_mesh} in comparison with the PFEM and the SFEM over these polygonal meshes above. A noticeable point in this figure is the performances of the PFEM which tend to increase at higher refinement levels of the mesh. \textcolor{blue}{This demonstrates the disadvantage of the PFEM in adaptive mesh computations with the presence of hanging nodes which causes the singularity of shape function derivatives, discussed in \cite{Tabarraei_Sukumar_2007}}. Thus, it can be opined that the PCFEM is not only an excellent numerical method in terms of both yielding accurate results and to treat adaptive meshes which will be necessary for the damage models involving steep gradients. 

\section{Formulations of nonlocal damage models}
\label{sec damage_formulations}

In this section, the continuum damage models based on the nonlocal integral formulation is established within the framework of PCFEM with a low-order approximation of the assumed strain field. The nonlocal integral is applied to compute the nonlocal equivalent strain that is estimated by weighting and averaging distributions of the local strain within an internal size.
\subsection{Fundamentals of continuum damage model}
\label{subsec damage_model}
As regards the physical aspect, the damage in solids arises from microstructural defects of materials, where a scalar variable, named as the damage variable is introduced to express the level of damage. Considering the damage variable, the constitutive relation between the strain-stress is given by:
\begin{equation}
          {\boldsymbol{\sigma }} = \left( {1 - \omega } \right){\bf{C}}:{\boldsymbol{\varepsilon }}
          \label{Eq.Consti_rela}
\end{equation}
where $\mathbf{C}$ is the fourth-order elastic material tensor; $\boldsymbol{\sigma}$ and $\boldsymbol{\varepsilon }$ are the stress and strain tensor, respectively. The value of the damage variable $\omega$ varies from $0$ to $1$, i.e, $0$ for intact material whereas $1$ for complete failure.\\
The variable $\omega$ is the damage evolution function of a state scalar variable $\kappa$ which tends to only monotonically increase. The evolution is governed by the Kuhn-Tucker condition as follows:
\begin{equation}
    \left\{ {\begin{array}{*{20}{c}}
	{f \le 0}\\
	{\dot \kappa  \ge 0}\\
	{f\dot \kappa  = 0}
	\end{array}} \right.
    \label{Eq.Kuhr-Tucker}
\end{equation}
where $f\left( {{\varepsilon _{eq}},\kappa } \right) = {\varepsilon _{eq}}\left( {\boldsymbol{\varepsilon }} \right) - \kappa$ is the damage loading function. From the above constraints, the value of the equivalent strain $\varepsilon _{eq}$ is actually smaller or equal to the so-called threshold equivalent strain $\kappa$, which corresponds to the maximum value of $\varepsilon _{eq}$ experienced by the material in the previous load step. \\ 

The equivalent strain is defined by the two common criteria, presented in: \cite{Peerlings_Borst_Brekelmans_1998} as follows
\begin{itemize}
	\item Mazars
	\begin{equation}
	    {\varepsilon _{eq}} = \sqrt {\sum\limits_{I = 1}^3 {{{\left( {\left\langle {{\varepsilon _I}} \right\rangle } \right)}^2}} }
	    \label{Eq.Mazars}
	\end{equation}
	where $\varepsilon _{I}$ are the principal strains, and the $\left\langle \cdot \right\rangle $ represents a positive operator, i.e.,$\left\langle \varepsilon  \right\rangle  = \varepsilon$ if $\varepsilon  \ge 0$, and $\left\langle \varepsilon  \right\rangle  = 0$ if $\varepsilon  < 0$.\\
	The measure is more appropriate to damage models under tensile conditions than compression, because negative strains are eliminated. Therefore, the damage growth is mainly caused by mode $I$ fracture.
	\item Modified von Mises
	\begin{equation}
    	{\varepsilon _{eq}} = \frac{{k - 1}}{{2k\left( {1 - 2\nu } \right)}}{I_1} + \frac{1}{{2k}}\sqrt {\frac{{{{\left( {k - 1} \right)}^2}}}{{{{\left( {1 - 2\nu } \right)}^2}}}I_1^2 + \frac{{12k}}{{{{\left( {1 + \nu } \right)}^2}}}J_2^{'}}
    	\label{Eq.vonMises}
	\end{equation}
	in which $k$ is a parameter corresponding to the ratio of the compressive to the tensile strength of the material, and $\nu$ is the Poisson's ratio. The terms $I_1$ and $J_2^{'}$ are the strain invariants given by:
	\begin{equation}
    	{I_1} = tr\left( {\boldsymbol{\varepsilon }} \right)
	\end{equation}
	\begin{equation}
        J_2^{'} = \frac{1}{6}\left( {3 tr\left( {{\boldsymbol{\varepsilon }} \cdot {\boldsymbol{\varepsilon }}} \right) - t{r^2}\left( {\boldsymbol{\varepsilon }} \right)} \right)
	\end{equation}
\end{itemize}

An exponential function proposed by Peerlings \cite{Peerlings_Borst_Brekelmans_1998} is introduced to describe the softening stress-strain relation as
\begin{equation}
     \omega \left( \kappa  \right) = \left\{ {\begin{array}{*{20}{c}}
	0&{{\rm{if }\,}\kappa  \le {\kappa _0}}\\
	{1 - \dfrac{{{\kappa _0}}}{\kappa }\left( {1 - \alpha  + \alpha \exp \left( { - \beta \left( {\kappa  - {\kappa _0}} \right)} \right)} \right)}&{{\rm{otherwise}}}
	\end{array}} \right.
      \label{Eq.damage_func}
\end{equation}
where $\alpha$ and $\beta$ are the material parameters, and $\kappa_0$ is the initial damage threshold value. The damage evolution is activated or the material is transferred to softening when values of the state variable $\kappa$ of the equivalent strain exceeds the initial threshold value $\kappa_0$. The parameter $\beta$ determines the speed of the damage growth.

\subsection{Nonlocal integral formulation}
\label{subsec: nonlocal integral}
The aim of using the nonlocal integral model for evaluating the damage variable is to ease the numerical hassles from the local formulations. Early work on this technique was applied into elastoplasticity \cite{Aifantis_1984}, damage \cite{Bazant_Lin_1988}, plasticity \cite{Bazant_Jirasek_2002} to restrict the regularizing effect resulting from the localized deformations. The underlying idea for damage models is a replacement of the equivalent strain $\varepsilon _{eq}$ at a point by a nonlocal counterpart  $\bar \varepsilon _{eq}$. The evaluation of the nonlocal equivalent strain is based on the weighted average of the local quantities distributed over neighbours of a point under consideration, bounded in a domain $V$. The mathematical expression is given by:
\begin{equation}
    {\bar \varepsilon _{eq}}\left( {{{\bf{x}}^p}} \right) = \int_V {\alpha \left( {{\bf{x}},{{\bf{x}}^p}} \right)} {\varepsilon _{eq}}\left( {\bf{x}} \right)d{\bf{x}}
    \label{Eq_eqbar}
\end{equation}
where $\alpha \left( {{\bf{x}},{{\bf{x}}^p}} \right)$ is an assumed nonlocal weight function. The definition is
\begin{equation}
    \alpha \left( {{\bf{x}},{{\bf{x}}^p}} \right) = \frac{{{\alpha _0}\left( {\left\| {{\bf{x}} - {{\bf{x}}^p}} \right\|} \right)}}{{\int_V {{\alpha _0}\left( {\left\| {{\bf{x}} - {{\bf{x}}^p}} \right\|} \right)d{\bf{x}}} }}
    \label{Eq.nonlocal_func}
\end{equation}
where ${\alpha _0}\left( {\left\| {{\bf{x}} - {{\bf{x}}^p}} \right\|} \right)$ is a weight function, also known as a kernel function is a monotonically non-negative  function of the distance $r = \left\| {{\bf{x}} - {{\bf{x}}^p}} \right\|$ which is defined between the consideration point ${{\bf{x}}^p}$ and the neighbour points $\bf{x}$. Normally, it is formed by the Gauss distribution function
\begin{equation}
    {\alpha _0}\left( r \right) = \exp \left( { - \frac{{{r^2}}}{{2l_c^2}}} \right)
    \label{Eq.Gauss_func}
\end{equation}
where $l_c$ is the internal length, also known as the nonlocal characteristic length scale. It depends on material properties such as the size of heterogeneous materials or the specification of failure mechanism. The kernel function given in Eq. \ref{Eq.Gauss_func} which directly depends on the characteristic length $l_c$, has an unbounded definition on distribution points $\mathbf{x}$. This function could be imposed on the whole of body, but it reaches to zero if points move far from the point ${\mathbf{x}}^p$.\\

\begin{figure}[!httb]
	\centering
	\subfigure[]{
		\includegraphics[scale=1]{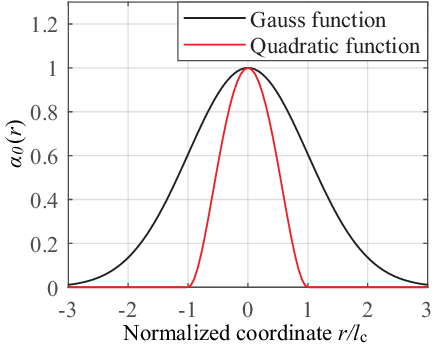}
		\label{Fig.kernel_1}
	} \hspace*{0.5em}
	\subfigure[]{
		\includegraphics[scale=1]{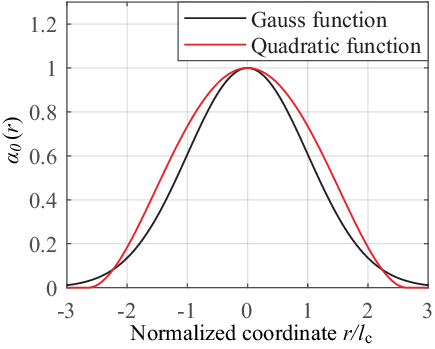}
		\label{Fig.kernel_sqrt7}
	}
	\caption{Weight functions corresponding to ratios of normalized coordinate: (a) $R = l_c$, (b) $R = \sqrt 7 {l_c}$.}
	\label{Fig.kernel_func}
\end{figure}

Another common kernel function is the truncated quadratic polynomial function, expressed as
\begin{equation}
    {\alpha _0}\left( r \right) = {\left\langle {1 - \frac{{{r^2}}}{{{R^2}}}} \right\rangle ^2}
    \label{Eq.Bel_func}
\end{equation}
where $R$ is a parameter related to the length scale $l_c$. The performances of kernel functions defined in Eqs. \ref{Eq.Gauss_func} \ref{Eq.Bel_func}, normalized with two different ratios $R = l_c$ and $R = \sqrt 7 {l_c}$ are described in Fig. \ref{Fig.kernel_func}.\\
Since the Gauss weight function is applied to the nonlocal formulation, a Taylor expansion of $\bar \varepsilon _{eq}$ from Eq. \ref{Eq.nonlocal_func} leads to form the nonlocal gradient formulation. The implicit gradient formulation could be a common approach in gradient-enhanced damage models \cite{Peerlings_Borst_Brekelmans_1998} and it is given by:
\begin{equation}
     {\bar \varepsilon _{eq}} - l_c^2{\nabla ^2}{\bar \varepsilon _{eq}} = {\varepsilon _{eq}}
     \label{Eq.Hemzholt}
\end{equation}

\begin{figure}[!httb]
	\centering
	\subfigure[]{
		\includegraphics[scale=1]{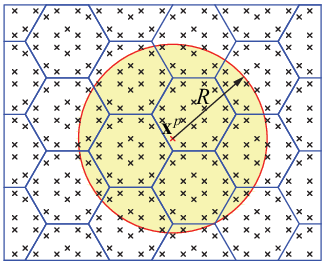}
		\label{Fig.kernel_mesh}
	} \hspace*{0.5em}
	\subfigure[]{
		\includegraphics[scale=1]{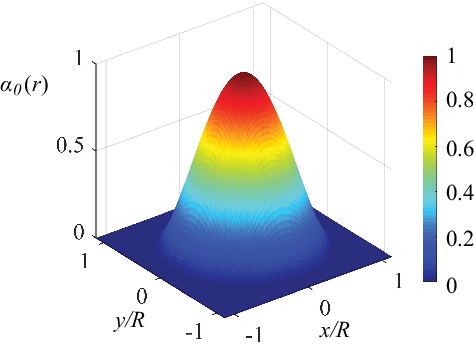}
		\label{Fig.kernel_Quad}
	}
	\caption{(a) Illustration of neighbor points of point ${\mathbf{x}}^p$, (b) the definition of the weight function in a polygonal mesh.}
	\label{Fig.kernel_mesh_quad}
\end{figure}

The second choice can be a close solution to well fit the description of the Gaussian distribution function, reported in Ref. \cite{Jirasek_2007}. Obviously, the distribution points within the window of influence, bounded by a circle of radius have significant effects on the weighting function. A description of the weight function defined in an extraction of a typically polygonal mesh is presented in Fig. \ref{Fig.kernel_mesh}, and the definition of this function with respect to normalized coordinates in two dimensions is plotted in Fig. \ref{Fig.kernel_mesh_quad}.\\
It is worth noticing that the mesh size at regions with the strong localization deformation should be sufficiently smaller than $R$. The purpose is to cover enough number of Gauss points within the effect of the kernel function with given interaction radius $R$.\\

To implement the nonlocal weighting function (c.f. Eq. \ref{Eq.nonlocal_func}) into the continuum damage formulation, we must numerically evaluate the weight of the nonlocal integral which are then associated with values of the local equivalent strain. The weight between two Gauss points $i$ and $j$ denoted by ${\alpha}_{ij}$ is computed as
 \begin{equation}
     {\alpha _{ij}} = \frac{{{\alpha _0}\left( {\left\| {{{\bf{x}}_i} - {{\bf{x}}_j}} \right\|} \right)}}{{\sum\limits_{m=1}^{n_{GP}^R} {{w_m}\left\| {{{\bf{J}}_m}} \right\|{\alpha _0}\left( {\left\| {{{\bf{x}}_i} - {{\bf{x}}_m}} \right\|} \right)} }}
     \label{Eq.nonlocal_num}
 \end{equation}
 in which $n_{GP}^R$ is the number of Gauss points located inside the domain of the interaction radius $R$, while $w$ and $\mathbf{J}$ are the integration weight and the Jacobian in the finite element model.
 The nonlocal equivalent strain at the integration point $i$ is accordingly defined as
 \begin{equation}
     {\bar \varepsilon _{eq,i}} = \sum\limits_{j=1}^{n_{GP}^R} {{w_j}\left\| {{{\bf{J}}_j}} \right\|{\alpha _{ij}}{\varepsilon _{eq,j}}}
     \label{Eq.nonlocal_strain_num}
 \end{equation}

 The computational process of the nonlocal approach practically takes a great deal of time due to searching neighbour integration points within the interaction domain centered at a certain integration point. Therefore, the tasks of determining the nonlocal integration points and the weight are normally executed in advance as \textit{an initialization procedure} to significantly decrease the computational cost. The steps involved are outlined as
 \begin{itemize}
 	\item Establishing positions of integration points over the whole computational domain.
 	\item For each integration point with a corresponding interaction domain of radius , seeking its neighbor integration points and calculating the coefficient ${a_{ij}} = {\alpha _0}\left( {\left\| {{{\bf{x}}_i} - {{\bf{x}}_j}} \right\|} \right){w_j}\left\| {{{\bf{J}}_j}} \right\|$, followed by computing the sum ${a_i} = \sum\limits_{j=1}^{n_{GP}^R} {{a_{ij}}}$.
 	\item Storing the individual information of integration point $i$ about $a_{ij}$ and $a_i$ into a data structure.
 \end{itemize}
Since the initialization procedure is done, computational tasks for the stiffness matrix accordingly carried out. Above all, the nonlocal equivalent strain ${\bar \varepsilon _{eq,i}}$ formulated in Eq. \ref{Eq.nonlocal_strain_num} can be rewritten by
\begin{equation}
     {\bar \varepsilon _{eq,i}} = \frac{{\sum\limits_{j = 1}^{n_{GP}^R} {{\varepsilon _{eq,j}}{a_{ij}}} }}{{{a_i}}}
     \label{Eq.nonlocal_strain_store}
\end{equation}
in which the coefficient $a_{ij}$ and the sum of them $a_i$ are directly taken from the data structure at point $i$. Thus, the task of determining neighbor points and their weight is recalled thanks to the built-in data during the iterative process of seeking numerical solutions.

\subsection{Weak form}
\label{subsec: weak form}

We consider a two-dimensional domain $\Omega$, bounded by the boundary $\Gamma$. The boundary conditions are given by ${\bf{t}} = {\bf{\bar t}}$ on the Neumann boundary ${\Gamma _t}$ and ${\bf{u}} = {\bf{\bar u}}$ on Dirichlet boundary ${\Gamma _u}$ such that ${\Gamma _t } \cap {\Gamma _u} = \emptyset $. The kinematic relation is assumed to be linear ${\boldsymbol{\varepsilon }} = {\nabla ^S}{\bf{u}}$, and the weak formulation is expressed as
\begin{equation}
    \int_\Omega  {\delta \nabla {\bf{u}}} :{\boldsymbol{\sigma }}d\Omega  = \int_{{\Gamma _t }} {\delta {{\bf{u}}^T} \cdot {\bf{\bar t}}d\Gamma }
    \label{Eq.weak1}
\end{equation}
Substituting the constitutive relation in Eq. \ref{Eq.Consti_rela} into Eq. \ref{Eq.weak1}, and the kinematic relation leads to the following form
\begin{equation}
    \int_\Omega  {\left( {1 - \omega } \right)\delta {{\boldsymbol{\varepsilon }}^T}}  : {\bf{C}} : {\boldsymbol{\varepsilon }}d\Omega  = \int_{{\Gamma _t}} {\delta {{\bf{u}}^T} \cdot {\bf{\bar t}}d\Gamma }
    \label{Eq.weak2}
\end{equation}
By applying the assumed strain method, discussed in sub-section. \ref{subsec: assumed strain}, the compatible strain field $\boldsymbol{\varepsilon }$ is replaced by $\boldsymbol{\tilde \varepsilon }$. Eq. \ref{Eq.weak2} is rewritten as
\begin{equation}
    \int_\Omega  {\left( {1 - \omega } \right)\delta {{{\boldsymbol{\tilde \varepsilon }}}^T}}  : {\bf{C}} : {\boldsymbol{\tilde \varepsilon }}d\Omega  = \int_{{\Gamma _t }} {\delta {{\bf{u}}^T} \cdot {\bf{\bar t}}d\Gamma }
    \label{Eq.weak3}
\end{equation}
The whole domain $\Omega$ is discretized into non-overlapping finite elements $\Omega_e$. The elemental displacement and strain fields are defined by means of shape functions $\boldsymbol{\phi}$ and the nodal displacements $\mathbf{d} _e$ with relations ${{\bf{u}}_e} = \boldsymbol{\phi} {{\bf{d}}_e}$ and ${\boldsymbol{\tilde \varepsilon }_e} = {{\bf{\tilde B}}_e}{{\bf{d}}_e}$, presented in Eq. \ref{Eq.assumed_strain}. Consequently, the weak form over each $\Omega_e$ is written by
\begin{equation}
    \delta {\bf{d}}_e^T\int_{{\Omega _e}} {\left( {1 - \omega } \right){{{\bf{\tilde B}}}_e}^T{\bf{C}}{{{\bf{\tilde B}}}_e}{{\bf{d}}_e}d\Omega }  = \delta {\bf{d}}_e^T\int_{\Gamma _t ^e} {\boldsymbol{\phi} ^T {\bf{\bar t}}d\Gamma }
    \label{Eq.weak4}
\end{equation}
The above equation, represented for the equilibrium of the discretization equation system can be expressed as
\begin{equation}
    {{\bf{f}}_{{\mathop{\rm int}} }}\left( {\bf{d}} \right) = {{\bf{f}}_{ext}}
    \label{Eq.equilibrium}
\end{equation}
where $\mathbf{f}_{int}$ is the internal force vector
\begin{equation}
    {{\bf{f}}_{{\mathop{\rm int}} }}\left( {\bf{d}} \right) = \int_\Omega  {\left( {1 - \omega } \right){{{\bf{\tilde B}}}^T}{\bf{C\tilde Bd}}d\Omega }
    \label{Eq.int_force}
\end{equation}
and $\mathbf{f}_{ext}$ is the external force vector
\begin{equation}
    {{\bf{f}}_{ext}} = \int_{{\Gamma _t }} {{\boldsymbol{\phi}^T}{\bf{\bar t}}d\Gamma }
    \label{Eq.ext_force}
\end{equation}
Obviously, Eq. \ref{Eq.equilibrium} is highly nonlinear. Thus, the linearization of this equation must be carried out before the iterative procedure is applied to seek solutions. The linearized relation can be defined as
\begin{equation}
    {{\bf{K}}_{\tan }}\Delta {\bf{d}} = {{\bf{f}}_{{\mathop{\rm int}} }}\left( {\bf{d}} \right) - {{\bf{f}}_{ext}}
    \label{Eq.linear_eq}
\end{equation}
Unlike the conventional nonlinear finite models, the tangent stiffness matrix of an element in nonlocal models is comprised of many blocks of matrix. Each block is mathematically defined by the derivative of the internal force vector of element $o$, denoted by ${\bf{f}}_{{\mathop{\rm int}} }^{\left( o \right)}$ with respect to the displacement vector ${{\bf{d}}^{\left( q \right)}}$ of element $q$ within the nonlocal interaction domain. The expression is
\begin{equation}
    \label{Eq.deri_fint}
    \begin{multlined}
        {\bf{K}}_{\tan }^{\left( {o,q} \right)} = \dfrac{{\partial {\bf{f}}_{{\mathop{\rm int}} }^{\left( o \right)}}}{{\partial {{\bf{d}}^{\left( q \right)}}}} = \dfrac{\partial }{{\partial {{\bf{d}}^{\left( q \right)}}}} {\int_{{\Omega ^{\left( o \right)}}} {\left( {1 - \omega } \right) {\bf{\tilde B}} ^{{\left( o \right)}T}} {\mathbf{C}}  {\bf{\tilde B}} ^{{\left( o \right)}} {{\bf{d}} ^{{\left( o \right)}}} d\Omega} \vspace{0.3cm}\\
        \quad \;\ =  {\int_{{\Omega ^{\left( o \right)}}} {\left( {1 - \omega } \right)} {{\bf{\tilde B}} ^{{\left( o \right)}T}} {\mathbf{C}}  {\bf{\tilde B}} ^{{\left( o \right)}}  \dfrac{\partial {{\bf{d}} ^{{\left( o \right)}}}}{{\partial {{\bf{d}}^{\left( q \right)}}}} d\Omega} -  {\int_{{\Omega ^{\left( o \right)}}} \dfrac{\partial {\omega}}{{\partial {{\bf{d}}^{\left( q \right)}}}} {{\bf{\tilde B}} ^{{\left( o \right)}T}} {\mathbf{C}}  {\bf{\tilde B}} ^{{\left( o \right)}} {{\bf{d}} ^{{\left( o \right)}}} d\Omega}
    \end{multlined}
\end{equation}
In Eq. \ref{Eq.deri_fint}, components in the first term, the so-called local stiffness matrix are valid with $o \equiv q$, and they are zeros in otherwise. Its definition is given by
\begin{equation}
    \label{Eq.local}
    \begin{multlined}
        K_l^{\left( {o,q} \right)} = \left\{ {\begin{array}{*{20}{c}}
    	{\bf{0}} \vspace{0.3cm}\\
    	{{\int_{{\Omega ^{\left( o \right)}}} {\left( {1 - \omega } \right)} {{\bf{\tilde B}} ^{{\left( o \right)}T}} {\mathbf{C}}  {\bf{\tilde B}} ^{{\left( o \right)}}  \dfrac{\partial {{\bf{d}} ^{{\left( o \right)}}}}{{\partial {{\bf{d}}^{\left( q \right)}}}} d\Omega} }
    	\end{array}} \right.  \vspace{0.3cm}\\
        \quad \;\ = \left\{ {\begin{array}{*{20}{c}}
        	{\bf{0}}&{{\rm{if }\,}o \ne q} \vspace{0.3cm}\\
        	{\sum\limits_i^{{n_{GP}}} {\left( {1 - {\omega _i}} \right){\bf{\tilde B}}{{_i^{\left( o \right) T}}}{\bf{C\tilde B}}_i^{\left( o \right)}{w_i}\left\| {{{\bf{J}}_i}} \right\|} }&{{\rm{if }\,}o \equiv q}
        	\end{array}} \right.
    \end{multlined}
\end{equation}
and the second term, the so-called nonlocal stiffness matrix is
\begin{equation}
    \begin{multlined}
        K_n ^{\left( {o,q} \right)} = {\int_{{\Omega ^{\left( o \right)}}} \dfrac{\partial {\omega}}{{\partial {{\bf{d}}^{\left( q \right)}}}} {{\bf{\tilde B}} ^{{\left( o \right)}T}} {\mathbf{C}}  {\bf{\tilde B}} ^{{\left( o \right)}} {{\bf{d}} ^{{\left( o \right)}}} d\Omega} \vspace{0.3cm}\\
        \qquad \;\ = {\int_{{\Omega ^{\left( o \right)}}} {\dfrac{{\partial {\omega ^{\left( o \right)}}}}{{\partial {\kappa ^{\left( o \right)}}}}} {\dfrac{{\partial {\kappa ^{\left( o \right)}}}}{{\partial \bar \varepsilon _{eq}^{\left( o \right)}}}} {\dfrac{{\partial \bar \varepsilon _{eq}^{\left( o \right)}}}{{\partial \varepsilon _{eq}^{\left( q \right)}}}} {\dfrac{{\partial \varepsilon _{eq}^{\left( q \right)}}}{{\partial {{\boldsymbol{\tilde \varepsilon }}^{\left( q \right)}}}}} {\dfrac{{\partial {{\boldsymbol{\tilde \varepsilon }}^{\left( q \right)}}}}{{\partial {{\bf{d}}^{\left( q \right)}}}}} {{\bf{\tilde B}}^{\left( o \right) T}} {\left( {{\bf{C}}{{{\bf{\tilde B}}}^{\left( o \right)}}{{\bf{d}}^{\left( o \right)}}} \right)} d\Omega } \vspace{0.3cm}\\
        =\sum\limits_i^{{n_{GP}}} {\sum\limits_j^{{n_{GP}^R}} {{\omega '} _i^{\left( o \right)}} {\dfrac{{\partial {\kappa ^{\left( o \right)}}}}{{\partial \bar \varepsilon _{eq}^{\left( o \right)}}}} {a_{ij}}w_i^{^{\left( o \right)}}\left\| {{\bf{J}}_i^{\left( o \right)}} \right\| {\bf{\tilde B}}{{_i^{\left( o \right)}}^T}{\boldsymbol{\sigma }}_i^{\left( o \right)}{\boldsymbol{\eta }}_j^{\left( q \right)T} {\bf{\tilde B}}_j^{\left( q \right)} }
    \end{multlined}
    \label{Eq.nonlocal}
\end{equation}
in which terms $\omega '$ and $\boldsymbol{\eta }$ are respectively denoted for the relation $\omega ' = \dfrac{{\partial \omega }}{{\partial \kappa }}$ and ${\boldsymbol{\eta }} = \dfrac{{\partial {\varepsilon _{eq}}}}{{\partial {\boldsymbol{\tilde \varepsilon }}}}$ that are specifically reported in \ref{sec:appendix}. The term $\dfrac{{\partial \kappa }}{{\partial {{\bar \varepsilon }_{eq}}}}$ is set to $1$ for loading and is equal to $0$ for unloading, while $a_{ij}$ is the coefficient of nonlocal interaction between integration points $i$ and $j$ defined in sub-section \ref{subsec: nonlocal integral}. Herein, $j$ is a set of integration points inside the interaction domain centered at the integration point $i$.\\
As seen that the nonlocal stiffness matrix expressed in Eq. \ref{Eq.nonlocal} that is valid in the loading regime is represented for the damage growth while the local counterpart shown in Eq. \ref{Eq.local} is considered as the stiffness of matrix in the unloading case. It should be noted that the consequence of double sums in Eq. \ref{Eq.nonlocal} is carried out for each pair of integration point $i$ with set of contributed integration points $j$ and then is assembled into the global tangent stiffness matrix.

\section{Numerical examples}
\label{sec: Numerical examp}
In this section, a set of three benchmark problems are designed to examine the mechanical behaviours of typical mode and mixed-mode fracture processes. The objective is to verify the computational efficiency of the present numerical method (PCFEM) in nonlocal damage mechanics. Unless mentioned otherwise, brittle materials are chosen and plane stress conditions are considered for simulations.

\subsection{Three-point bending beam test}
\label{subsec: Three-point bending beam test}
\begin{figure}[!httb]
	\centering
	\includegraphics[scale=1]{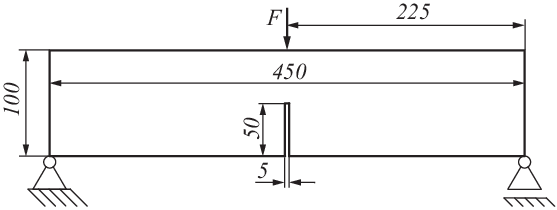}
	\caption{Geometry and boundary conditions of a beam with a notch.}
	\label{Fig.notched_geo}
\end{figure}

\begin{figure}[!httb]
	\centering
	\includegraphics[scale=1]{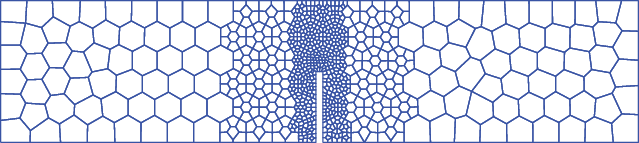}
	\caption{Polygonal mesh of the notched beam.}
	\label{Fig.notched_mesh}
\end{figure}

This example investigates the damage in mode I in a concrete beam with a pre-exissting notch. An experiment for this model was conducted by Kormeling
\cite{Kormeling_1983}, and several numerical simulations were presented by Jir\'asek \cite{Jirasek_2007}, Bobinski \cite{Bobinski_Tejchman_2005}, Lorentz \cite{Lorentz_2017}. As a good example for the nonlocal damage model, Jir\'asek's results are taken into comparison with performances of the present numerical method. According to his report, the beam has thickness $t = $ 100 mm, and its other geometrical parameters are shown in Fig. \ref{Fig.notched_geo}. The material parameters of concrete are chosen as Young's modulus $E = $20 GPa and Poisson's ratio $\nu =$ 0.2, and the equivalent strain is measured based on Mazars criterion given in Eq. \ref{Eq.Mazars}. The damage parameters adopted to Eq. \ref{Eq.damage_func} are assumed to be $\alpha = $ 0.98, $\beta =$ 300, and the limiting elastic strain $\kappa _0 =$ 90$\times $10$^{ - 6}$. The interaction radius is set to $R = $ 4 mm. The discretization of the domain is carried out with a polygonal mesh equipped with local refinements that cover the damage zone, as shown in Fig. \ref{Fig.notched_mesh}. \textcolor{blue}{Three meshes corresponding to minimum mesh sizes $6.55\ mm$, $2.57\ mm$ and $2.01\ mm$ are used for the analysis}. The vertical displacement at the point where the external force $F$ is observed.\\

\begin{figure}[!httb]
	\centering
	\includegraphics[scale=1]{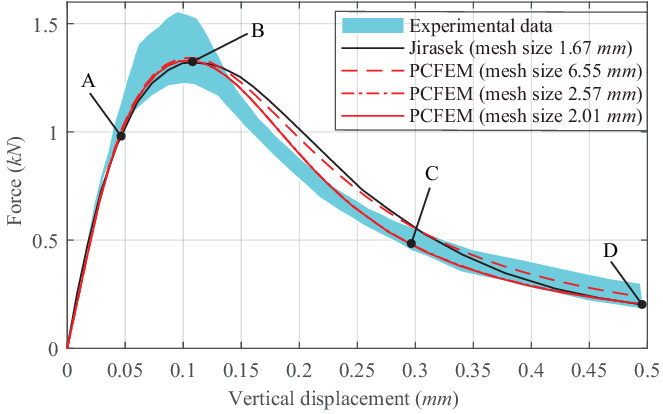}
	\caption{\textcolor{blue}{Force-displacement curves of the notched beam.}}
	\label{Fig.notched_disp_force}
\end{figure}

\begin{figure}[!httb]
	\centering
	\includegraphics[scale=1]{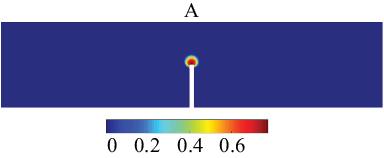} \hspace{0.5cm}
	\includegraphics[scale=1]{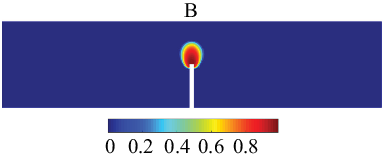}\\
	\includegraphics[scale=1]{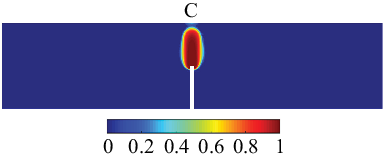}\hspace{0.5cm}
	\includegraphics[scale=1]{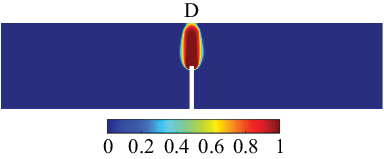}
	\caption{Evolution of the damage profile of the notched beam.}
	\label{Fig.notched_damge}
\end{figure}

The force-displacement relationship, identified as the damage behavior of this model is depicted in Fig. \ref{Fig.notched_disp_force}. \textcolor{blue}{As observed, the result from the mesh size of $2.57\ mm$ is almost identical to that of $2.01\ mm$, and they look different from the performance of the mesh size of $6.55\ mm$. The curve corresponding to the coarse mesh exhibits higher values in the softening region. It could be confirmed that the two finer meshes provide the converged solution when their sizes are smaller than the interaction length. Whereas the size of the coarse mesh is bigger compared with the interaction length, this lead to insufficient accuracy in resolving the localized process zone. Back to Fig. \ref{Fig.notched_disp_force}, the PCFEM produces the converged solutions with excellent agreement with the experimental data \cite{Jirasek_2007} during almost the loading history. In comparison with Jir\'asek's result, these solutions look better in capturing the softening response}. The evolution of the damage profile with corresponding states marked in Fig. \ref{Fig.notched_disp_force} is plotted in Fig. \ref{Fig.notched_damge}. As can be seen in this figure, the damage zone is concentrated and grow along the axis of symmetry of the structure.\\

\begin{figure}[!httb]
	\centering
	\includegraphics[scale=1]{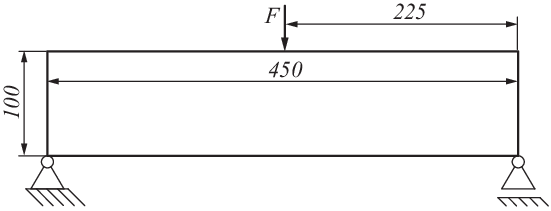}
	\caption{Geometry and boundary conditions of a beam without a notch.}
	\label{Fig.unnotched_geo}
\end{figure}

\begin{figure}[!httb]
	\centering
	\includegraphics[scale=1]{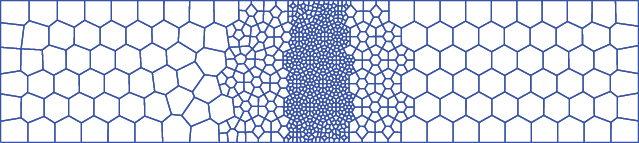}
	\caption{Polygonal mesh of the unnotched beam.}
	\label{Fig.unnotched_mesh}
\end{figure}

Next, the three-point bending test is considered for the case of the beam without a notch. As shown in Jir\'asek's report \cite{Jirasek_2007}, the material parameters, and limit elastic strain are reused as same as the notched beam. The size of the beam is kept for the examination as described in Fig. \ref{Fig.unnotched_geo}, and a polygonal mesh with refinements is shown in Fig. \ref{Fig.unnotched_mesh}. Only change is the nonlocal interaction radius $R =$ 8 mm.\\

\begin{figure}[!httb]
	\centering
	\includegraphics[scale=1]{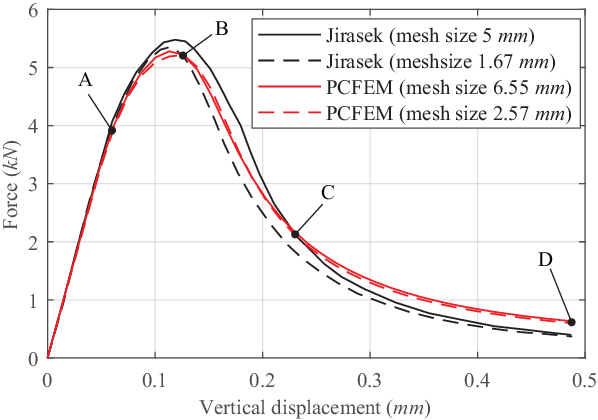}
	\caption{Force-displacement curves of the unnotched beam.}
	\label{Fig.unnotched_disp_force}
\end{figure}

\begin{figure}[!httb]
	\centering
	\includegraphics[scale=1]{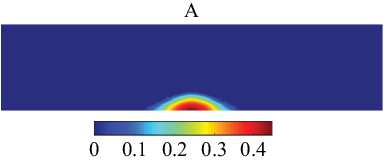} \hspace{0.5cm}
	\includegraphics[scale=1]{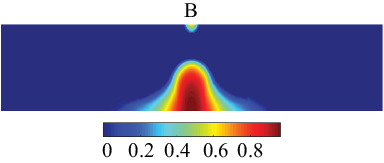}\\
	\includegraphics[scale=1]{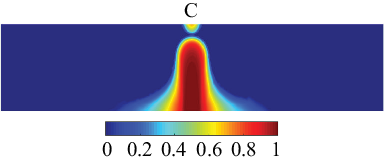}\hspace{0.5cm}
	\includegraphics[scale=1]{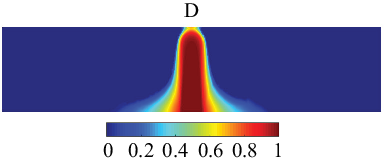}
	\caption{Evolution of the damage profile of the unnotched beam.}
	\label{Fig.unnotched_damge}
\end{figure}

\textcolor{blue}{The test is analyzed on two meshes with the sizes of $6.55\ mm$ and $2.57\ mm$, respectively. The force-displacement relations corresponding to the two meshes are plotted in Fig. \ref{Fig.unnotched_disp_force}}. It can be inferred that there is no difference in the force-displacement curves of the present method (PCFEM) between the two minimum mesh sizes which are chosen to be smaller than the interaction length. Therefore, the selection of the local mesh sizes must be sufficient enough to capture the damage zone where localized strain is highly activated. The PCFEM provides accurate solutions that are favorably compared with former numerical results obtained from the finite element framework \cite{Jirasek_2007}. Some of the damage stages in the damage process marked in Fig. \ref{Fig.unnotched_disp_force} are plotted in Fig. \ref{Fig.unnotched_damge}.\\

\begin{figure}[!httb]
	\centering
	\includegraphics[scale=1]{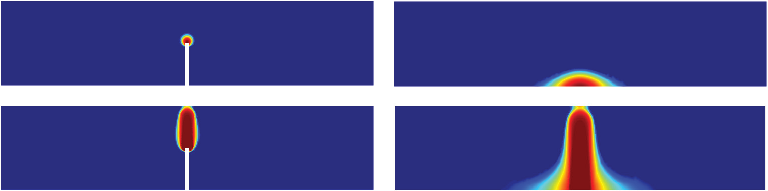}
	\caption{\textcolor{blue}{Comparison of the damage profile of the notched and unnotched beam.}}
	\label{Fig.damage_notched_unnotched}
\end{figure}

According to the above numerical examples for the three-point bending test of a notched and unnotched beam, the nonlocal model is of importance for damage problems because of the existence of the pathological mesh-sensitivity caused by localized deformations. It is seen that if the mesh size is sufficiently fine, or in other words it must be smaller than the interaction length, the obtained results become independent on the mesh density. It can be noted that the choice of the interaction length must be consistent with the band width of the damage zone where highly localized deformations are distribute. \textcolor{blue}{Looking at distributions of the damage process zones shown in Fig. \ref{Fig.damage_notched_unnotched}, the narrow band in the unnotched beam is larger than the notched counterpart, and this is the reason why the interaction length corresponding to the unnotched beam was chosen to be bigger.}

\subsection{L-shaped concrete specimen}
\label{subsec: L-shaped concrete specimen}
\begin{figure}[!httb]
	\centering
	\includegraphics[scale=1]{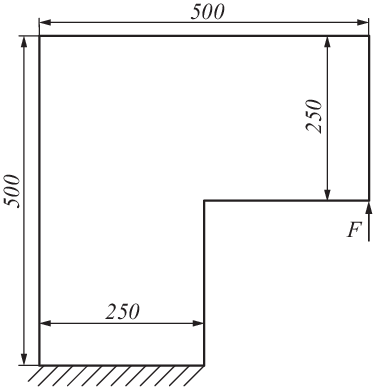}
	\caption{Geometry and boundary conditions of the L-shaped specimen.}
	\label{Fig.Lshape_geo}
\end{figure}

\begin{figure}[!httb]
	\centering
	\includegraphics[scale=1]{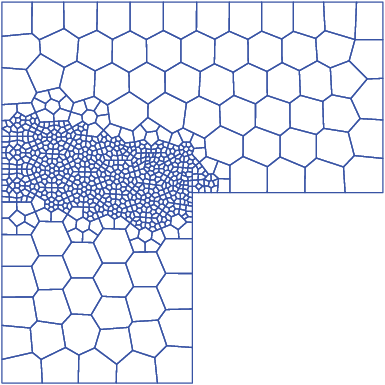}
	\caption{Polygonal mesh of the L-shaped specimen.}
	\label{Fig.Lshape_mesh}
\end{figure}

This example is to investigate the damage behaviours in mixed-mode fracture. The benchmark was introduced by Winkler \cite{Winkler_Hofstetter_Niederwanger_2001} in experimental tests for concrete cracking under smeared crack approach. The geometry of the specimen with thickness $t =$ 100
mm is described in Fig. \ref{Fig.Lshape_geo}, where the short-bottom edge is clamped and a concentrated load is imposed to the corner of two short edges at the right hand side. The material parameters of concrete are given as Young’s modulus $E=$ 25850 MPa, Poisson’s ratio $\nu=$ 0.18 and tensile strength $\sigma_t =$ 2.7 MPa and with an interaction length $R=$ 10 mm. Due to the effect of mixed-mode loading, the von Mises criterion in Eq. \ref{Eq.vonMises} is adopted to analysis with the ratio of tensile and compressive strength $k=$ 10. The damage parameters are chosen as $\alpha =$ 0.98 and $\beta =$ 350, while the initial threshold value $\kappa_0$ is taken to be the ratio $\dfrac{{{\sigma _t}}}{E}$ presented in Ref. \cite{Winkler_Hofstetter_Niederwanger_2001}. A polygonal mesh with $760$ elements as shown in Fig \ref{Fig.Lshape_mesh}. is employed to numerically analyze, and the vertical displacement is observed at the point where the force is applied.\\

\begin{figure}[!httb]
	\centering
	\includegraphics[scale=1]{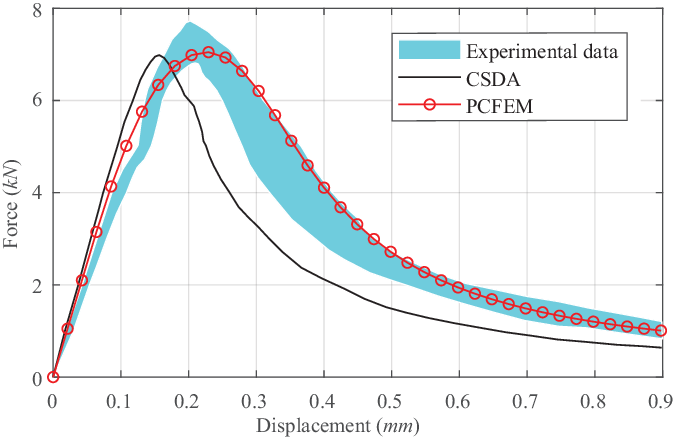}
	\caption{Force-displacement curves of the L-shaped specimen.}
	\label{Fig.Lshape_disp_force}
\end{figure}

Fig. \ref{Fig.Lshape_disp_force} shows the displacement-force curves of this model in which a published numerical result obtained from the Continuum Strong Discontinuity Approach (CSDA) with 6645 triangular elements, reported in \cite{Oliver_Huespe_Pulido_2004} is taken for comparison. Herein, the performance of the CSDA has the same configuration to the experimental feature, but it does not fall in the distribution of experimental data. While the present numerical method (PCFEM) offers an excellent solution almost located inside the experimental observation only with $760$ elements.\\

\begin{figure}[!httb]
	\centering
	\subfigure[]{
		\includegraphics[scale=1]{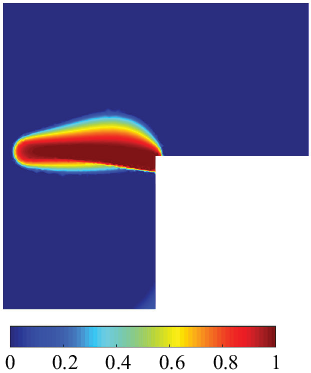}
		\label{Fig.Lshape_damage}
	} \hspace*{2em}
	\subfigure[]{
		\includegraphics[scale=1]{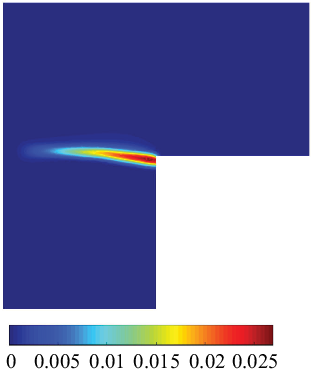}
		\label{Fig.Lshape_epseq}
	}\\
    \subfigure[]{
    	\includegraphics[scale=1]{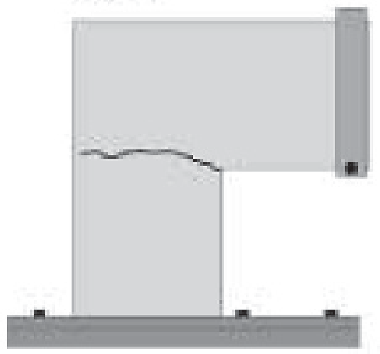}
    	\label{Fig.Lshape_exp}
    }
	\caption{(a) Damage profile, (b) equivalent strain of the L-shaped specimen, \textcolor{blue}{(c) crack path from the experiment \cite{Winkler_Hofstetter_Niederwanger_2001}}.}
	\label{Fig.Lshape_band}
\end{figure}

Fig. \ref{Fig.Lshape_damage} indicates the consequence of the damage zone that tends to be curved to upper side of the specimen. \textcolor{blue}{As a relevant proof of the damage process, the nonlocal equivalent strain distribution plotted in Fig. \ref{Fig.Lshape_epseq} is similarly configured to the damage profile where the material there is degraded. In addition, the direction of the damage evolution described by these two fields is well consistent with the experimental data shown in Fig. \ref{Fig.Lshape_exp}}.\\

\subsection{Mixed-mode fracture on a notched beam}
\label{subsec: Mixed-mode fracture on a notched beam}
\begin{figure}[!httb]
	\centering
	\includegraphics[scale=1]{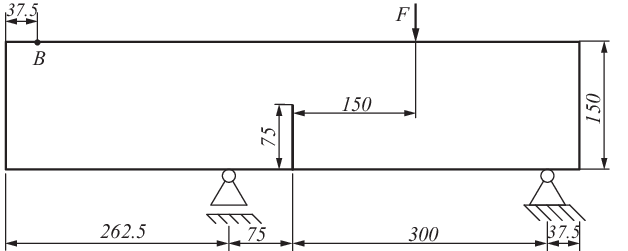}
	\caption{Geometry and boundary conditions for mixed-mode fracture of the beam.}
	\label{Fig.Galvez_geo}
\end{figure}

\begin{figure}[!httb]
	\centering
	\includegraphics[scale=1]{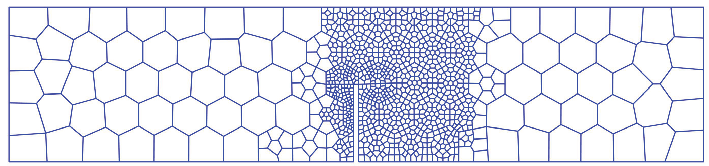}
	\caption{Polygonal mesh ($1014$ elements) for mixed-mode fracture of the beam.}
	\label{Fig.Galvez_mesh}
\end{figure}

\begin{figure}[!httb]
	\centering
	\includegraphics[scale=1]{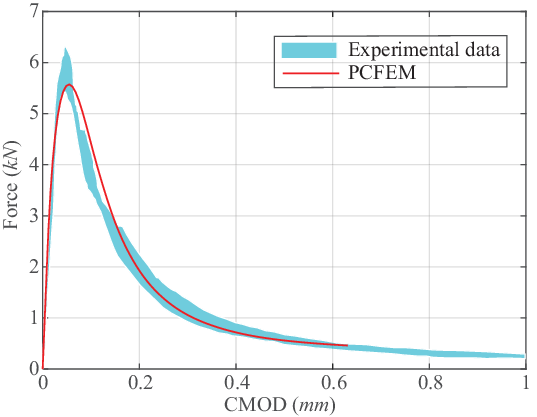}
	\caption{Curves of force $F$ and CMOD.}
	\label{Fig.disp_force_Galvez_CMOD}
\end{figure}

\begin{figure}[!httb]
	\centering
	\includegraphics[scale=1]{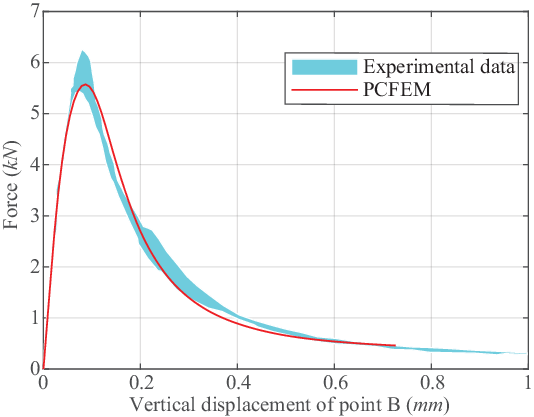}
	\caption{Curves of force $F$ and $B$}
	\label{Fig.disp_force_Galvez_B}
\end{figure}

\begin{figure}[!httb]
	\centering
	\subfigure[]{
		\includegraphics[scale=1]{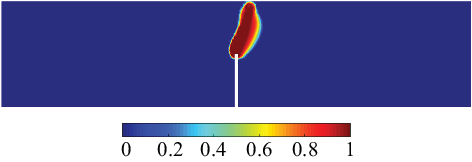}
		\label{Fig.Galvez_damage}
	} \hspace*{0.5em}
	\subfigure[]{
		\includegraphics[scale=1]{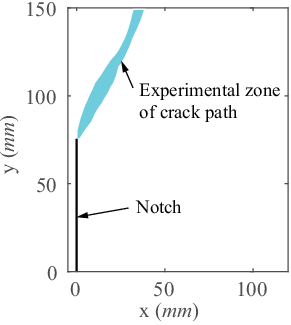}
		\label{Fig.Galvez_experiment}
	}
	\caption{(a) Damage profile for mixed-mode fracture of the beam, (b) experimental zone of crack path \cite{Galvez_Elices_Guinea_1998}.}
	\label{Fig.Galvez_damage_2}
\end{figure}

\begin{figure}[!httb]
	\centering
	\includegraphics[scale=1]{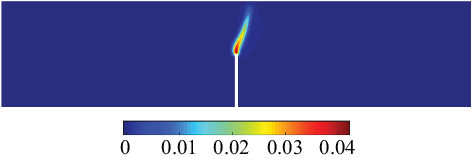}
	\caption{Equivalent strain for mixed-mode fracture of the beam.}
	\label{Fig.Galvez_strain}
\end{figure}

As a further example for the mixed-mode fracture, consider a beam with thickness $t =$ 50 mm as shown in Fig. \ref{Fig.Galvez_geo} with a notch 5 mm. The experimental results were reported by Galvez \cite{Galvez_Elices_Guinea_1998}  and this demonstrates the normal/shear cracking of concrete. The following material parameters are as Young’s modulus $E=$ 38 GPa, Poisson’s ratio $\nu=$ 0.18, tensile strength $\sigma_t =$ 3 MPa and compressive strength $\sigma_c =$ 57 MPa. The damage parameters are chosen as $\alpha =$ 0.98, $\beta =$ 400, and $\kappa_0 =$ 0.9. The interaction length is assumed to be $R=$ 5 mm. A mesh with $1014$ polygonal elements, shown in Fig. \ref{Fig.Galvez_mesh} is used for analysis. The vertical displacement at point $B$ and the crack mouth opening displacement (CMOD) are monitored. \\ 
Fig. \ref{Fig.disp_force_Galvez_CMOD} and Fig. \ref{Fig.disp_force_Galvez_B} show relationships between the force $F$ with CMOD and the vertical displacement of point $B$, respectively. As observed, the curves obtained from the present numerical method (PCFEM) agree quite well with the experimental results in both measuring approaches. Looking at these two figures, the stiff of the beam is significantly weak after reaching the peak load. This phenomenon happens more slowly in mode crack propagation, mentioned earlier in Example \ref{subsec: Three-point bending beam test} that has no effect of shear cracking. \\
The zone of the damage process in the final stage is indicated in Fig. \ref{Fig.Galvez_damage}. We can see that the direction of damage zone grows to the top edge and slopes to the right-hand side. Moreover, this numerical prediction shown in Fig. \ref{Fig.Galvez_experiment} coincides well with the layer of crack trajectories in the Galvez’s experiment \cite{Galvez_Elices_Guinea_1998}, attached in Fig. \ref{Fig.Galvez_experiment}. The corresponding equivalent strain distribution is plotted in Fig. \ref{Fig.Galvez_strain}, where a narrow band of highest values is tracked as same as the growth of the damage profile.\\

\subsection{\textcolor{blue}{Tensile test for a double-notched beam}}
\label{subsec:Tensile test for a double-notched beam}
\begin{figure}[!httb]
	\centering
	\subfigure[]{
		\includegraphics[scale=1]{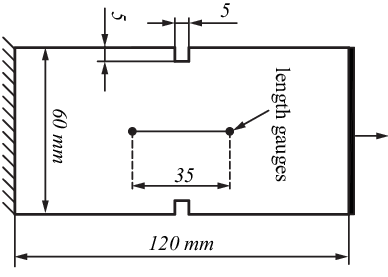}
		\label{Fig.double_notched_geo}
	} \hspace*{1.5em}
	\subfigure[]{
		\includegraphics[scale=0.9]{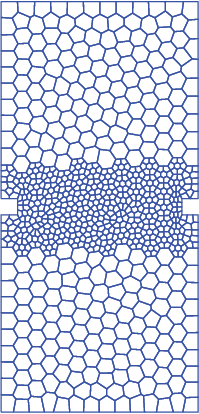}
		\label{Fig.doble_notched_mesh}
	}
	\caption{\textcolor{blue}{(a) Geometry and boundary conditions of double-notched specimen, (b) polygonal mesh (774 elements).}}
	\label{Fig.Geo_mesh_double_notched.}
\end{figure}

\textcolor{blue}{The example is designed for analyzing damage behaviors of multi-cracks in a concrete specimen. This model was first experimentally examined by Hordijk \cite{Hordijk_1991}, followed by a numerical investigation presented by Peerlings \cite{Peerlings_Borst_Brekelmans_1998} on various lightweight concrete specimens with double-edge notches. In this example, the chosen configuration of the model with a thickness of $50\ mm$ is shown in Fig. \ref{Fig.double_notched_geo} in which the bottom edge is fixed, and the top edge is pulled via a controlled displacement. A pair of gauges are installed to measure the vertical elongation $\delta$. Following Peerlings's report, material parameters are given by $E=18\ GPa$, $\nu=0.2$, and parameters for damage model are taken as $\alpha=0.96$, $\beta=350$, and equivalent strain is evaluated under the modified von-Mises criterion with $k=10$. The interaction length $R$ is set to $3\ mm$, and a mesh with $774$ elements shown in Fig. \ref{Fig.doble_notched_mesh} is taken into analysis.}\\

\begin{figure}[!httb]
	\centering
	\includegraphics[scale=1]{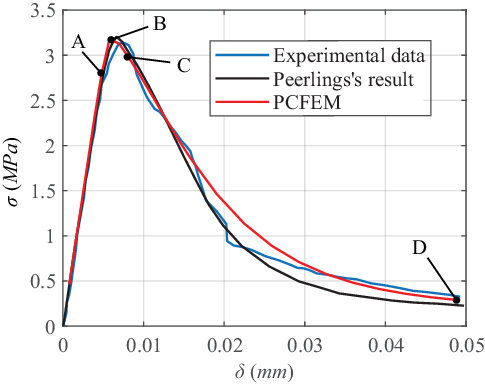}
	\caption{\textcolor{blue}{Stress-displacement curves for the damage process.}}
	\label{Fig.double_notched_curve}
\end{figure}

\textcolor{blue}{The first performance regarding structural behaviors is revealed by the stress-displacement diagrams shown in Fig. \ref{Fig.double_notched_curve} in which $\sigma$ is the stress with respect to the smallest cross-section of the specimen, and $\delta$ is the displacement obtained from the elongation of the $35\ mm$ distance. As shown in this figure, the consequence from PCFEM is almost consistent with data of the experiment \cite{Hordijk_1991}. In comparison with the former numerical result, PCFEM is only different from the softening stage with  a slightly higher stress.}

\begin{figure}[!httb]
	\centering
	\subfigure[]{
		\includegraphics[scale=1]{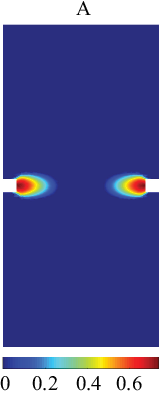}
	} \hspace*{0.5em}
	\subfigure[]{
		\includegraphics[scale=1]{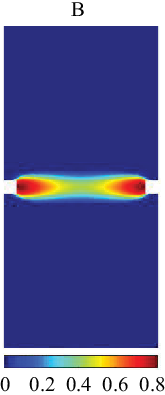}
	}\hspace*{0.5em}
    \subfigure[]{
    	\includegraphics[scale=1]{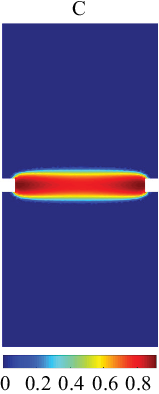}
    }\hspace*{0.5em}
    \subfigure[]{
    	\includegraphics[scale=1]{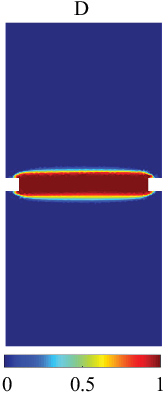}
    }
	\caption{\textcolor{blue}{Evolution of damage profile of the double-notched beam.}}
	\label{Fig.damage_profile_double_notched.}
\end{figure}

\textcolor{blue}{Fig. \ref{Fig.damage_profile_double_notched.} indicates the process of the damage growth in several stages marked in Fig. \ref{Fig.double_notched_curve}, and A, B and C are represented for first stages when the damage zones appear in the structure. As observed, the evolution of the damage profile propagates very quickly for a link-up of the damage zones from two notches. Moreover, the the growth of cracks tends to connect together and this is made for the crack intersection phenomenon due to their interaction mechanism.}

\section{Conclusions}
\label{sec:conclusions}

\textcolor{blue}{The study presented a further development of polygonal finite elements equipped with an assumed strain technique, named as \textit{polytopal composite finite elements} (PCFEM) to model problems of concrete fracture via the nonlocal damage models}. The application of the assumed strain technique on polygonal elements is to generate new strain fields which satisfies both the orthogonality condition, and the patch tests~\cite{Hung_2019}. In comparison with other numerical methods, the PCFEM provided higher-accurate solutions than the polygonal finite elements with rotational basis functions such as Wachspress functions, and the cell based SFEM. In addition, the PCFEM dealt with problems involving the less accuracy of derivatives of shape functions at points nearby the hanging nodes. This noticeable feature, validated in Section proved the advantage of the PCFEM in numerical simulations related to adaptive mesh computations. Therefore, formulations of the nonlocal damage model were established accordingly within the framework of PCFEM in which the nonlocal weight function was defined and the nonlocal tangent stiffness matrix was computed. The computational efficiency of the PCFEM was demonstrated through several numerical examples analyzed for typical cases of mode I to mixed-mode fracture processes. The PCFEM has yielded highly accurate solutions which are in good agreement with the experimental data.  Although, the nonlocal damage model coupled with the PCFEM offered the high accuracy of results, the computational time is cumbersome. Thus, the development of parallel computing is expected in the future work. On the other hand, the damage phenomenon goes along with various physical behaviour of the materials. The task of solving damage models with multi-fields will be interesting topics for the scientific communities of fracture. \\ 
\section*{Acknowledgments}
The support provided by RISE-project BESTOFRAC (734370)-H2020 is gratefully acknowledged.

%
%
%
%

\appendix
\section{Definition of derivatives of the damage evolution function and the equivalent strain}
\label{sec:appendix}

\textit{Derivative of the damage evolution function} $\omega ' = \dfrac{{\partial \omega }}{{\partial \kappa }}$
\begin{equation}
    \label{Eq.deri_damage_func}
    \begin{multlined}
        \omega ' = \dfrac{{\partial \omega }}{{\partial \kappa }} = \left\{ {\begin{array}{*{20}{c}}
    	0&{{\rm{if }\,}\kappa  \le {\kappa _0}} \vspace{0.3cm}\\
	{1 - \dfrac{{{\kappa _0}}}{\kappa }\left( {1 - \alpha  + \alpha \exp \left( { - \beta \left( {\kappa  - {\kappa _0}} \right)} \right)} \right)}&{{\rm{otherwise}}}
	\end{array}} \right. \vspace{0.3cm}\\
{\rm{  }} = \left\{ {\begin{array}{*{20}{c}}
	0&{{\rm{if }\,}\kappa  \le {\kappa _0}} \vspace{0.3cm}\\
	{\dfrac{{{\kappa _0}}}{{{\kappa ^2}}}\left( {1 - \alpha  + \alpha \exp \left( { - \beta \left( {\kappa  - {\kappa _0}} \right)} \right)} \right) + \dfrac{{{\kappa _0}}}{\kappa }\alpha \beta \exp \left( { - \beta \left( {\kappa  - {\kappa _0}} \right)} \right)}&{{\rm{otherwise}}}
    	\end{array}} \right.
    \end{multlined}
\end{equation}

\textit{Derivatives of the equivalent strain with respect to the engineering strain tensor} ${\boldsymbol{\eta }} = \dfrac{{\partial {\varepsilon _{eq}}}}{{\partial {\boldsymbol{\tilde \varepsilon }}}}$\\

We first determine values of strain components and principal strain involving $z$-direction in the case of plane strain and plane stress condition.
\begin{itemize}
	\item For plane strain
	\begin{subequations}
		\begin{equation}
		    {\tilde \varepsilon _{xz}} = {\tilde \varepsilon _{yz}} = {\tilde \varepsilon _{zz}} = 0
		\end{equation}
		\begin{equation}
		    {\varepsilon _3} = 0
		\end{equation}
	\end{subequations}
    \item For plane stress
    \begin{subequations}
    	\begin{equation}
    	    {\tilde \varepsilon _{xz}} = {\tilde \varepsilon _{yz}} = 0, \,{\tilde \varepsilon _{zz}} =  - \dfrac{\nu }{{1 - \nu }}\left( {{\tilde \varepsilon _{xx}} + {\tilde \varepsilon _{yy}}} \right)
    	\end{equation}
    	\begin{equation}
        	{\varepsilon _3} = \tilde\varepsilon _{zz}
    	\end{equation}
    \end{subequations}
\end{itemize}

Derivatives of the equivalent strain computed by Mazars criterion in Eq. and modified von Mises criterion are expressed as
\begin{itemize}
	\item For Mazars criterion
	Principal strain in plane $xy$
	\begin{subequations}
		\begin{equation}
		    {\varepsilon _1} = a - b
		\end{equation}
		\begin{equation}
		{\varepsilon _1} = a + b
		\end{equation}
	\end{subequations}
    with $a = 0.5\left( {{\tilde \varepsilon _{xx}} + {\tilde \varepsilon _{yy}}} \right)$, and $b = \sqrt {0.5{{\left( {{\tilde \varepsilon _{xx}} - {\tilde \varepsilon _{yy}}} \right)}^2} + \tilde\varepsilon _{xy}^2}$.\\
    The non-negative part of the principal strains can be written by
    \begin{equation}
    \left\langle {{\varepsilon _I}} \right\rangle  = 0.5\left( {\left| {{\varepsilon _I}} \right| + {\varepsilon _I}} \right),{\rm{ }\:}I = 1,{\rm{ }}2,{\rm{ }}3
    \end{equation}
    and components of the principal strain are expressed under a tensor
    \begin{equation}
        {\bf{e}} = \left[ {\left\langle {{\varepsilon _1}} \right\rangle {\rm{ }}\left\langle {{\varepsilon _2}} \right\rangle {\rm{ }}\left\langle {{\varepsilon _3}} \right\rangle } \right]
    \end{equation}
    By applying the chain rule, we get
    \begin{equation}
        {\boldsymbol{\eta }} = \dfrac{{\partial {\varepsilon _{eq}}}}{{\partial {\boldsymbol{\tilde \varepsilon }}}} = \dfrac{{\partial {\varepsilon _{eq}}}}{{\partial {\bf{e}}}} \dfrac{{\partial {\bf{e}}}}{{\partial {\boldsymbol{\tilde \varepsilon }}}}
    \end{equation}
    in which $\dfrac{{\partial {\varepsilon _{eq}}}}{{\partial {\bf{e}}}} = \dfrac{1}{{{\varepsilon _{eq}}}}\left[ {\left\langle {{\varepsilon _1}} \right\rangle \ {\rm{ }}\left\langle {{\varepsilon _2}} \right\rangle \ {\rm{ }}\left\langle {{\varepsilon _3}} \right\rangle } \right]$, $\dfrac{{\partial {\bf{e}}}}{{\partial {\boldsymbol{\tilde \varepsilon }}}} = \left[ {\dfrac{{\partial \left\langle {{\varepsilon _1}} \right\rangle }}{{\partial {\boldsymbol{\tilde \varepsilon }}}} \ {\rm{ }}\dfrac{{\partial \left\langle {{\varepsilon _2}} \right\rangle }}{{\partial {\boldsymbol{\tilde \varepsilon }}}} \ {\rm{ }}\dfrac{{\partial \left\langle {{\varepsilon _3}} \right\rangle }}{{\partial {\boldsymbol{\tilde \varepsilon }}}}} \right]$.
    \item For modified von Mises criterion\\
    The equivalent strain in Eq. \ref{Eq.vonMises} could be written as
    \begin{equation}
        {\varepsilon _{eq}} = a{I_1} + b\sqrt {cI_1^2 + dJ_2^{'}}
    \end{equation}
    with $a = \dfrac{{k - 1}}{{2k\left( {1 - 2\nu } \right)}}$, $b = \dfrac{1}{{2k}}$, $c = \dfrac{{{{\left( {k - 1} \right)}^2}}}{{{{\left( {1 - 2\nu } \right)}^2}}}$, $d = \dfrac{{12k}}{{{{\left( {1 + \nu } \right)}^2}}}$.\\
    The derivatives of $\varepsilon_{eq}$ could be expressed
    \begin{equation}
        \dfrac{{\partial {\varepsilon _{eq}}}}{{\partial {\boldsymbol{\tilde \varepsilon }}}} = {\left[ {\dfrac{{\partial {\varepsilon _{eq}}}}{{\partial {\tilde\varepsilon _{xx}}}} \ {\rm{ }}\dfrac{{\partial {\varepsilon _{eq}}}}{{\partial {\tilde\varepsilon _{yy}}}} \ {\rm{ }}\dfrac{{\partial {\varepsilon _{eq}}}}{{\partial {\tilde \varepsilon _{xy}}}}} \right]^T}
    \end{equation}
    where\\
    \begin{equation}
        \dfrac{{\partial {\varepsilon _{eq}}}}{{\partial {\tilde\varepsilon _{xx}}}} = a\dfrac{{\partial {I_1}}}{{\partial {\tilde\varepsilon _{xx}}}} + \dfrac{{0.5b}}{{\sqrt {cI_1^2 + dJ_2^{'}} }}\left( {2c{I_1}\dfrac{{\partial {I_1}}}{{\partial {\tilde\varepsilon _{xx}}}} + d\dfrac{{\partial J_2^{'}}}{{\partial {\tilde\varepsilon _{xx}}}}} \right)
    \end{equation}
    \begin{equation}
        \dfrac{{\partial {\varepsilon _{eq}}}}{{\partial {\tilde\varepsilon _{yy}}}} = a\dfrac{{\partial {I_1}}}{{\partial {\tilde\varepsilon _{yy}}}} + \dfrac{{0.5b}}{{\sqrt {cI_1^2 + dJ_2^{'}} }}\left( {2c{I_1}\dfrac{{\partial {I_1}}}{{\partial {\tilde\varepsilon _{yy}}}} + d\dfrac{{\partial J_2^{'}}}{{\partial {\tilde\varepsilon _{yy}}}}} \right)
    \end{equation}
    \begin{equation}
        \dfrac{{\partial {\varepsilon _{eq}}}}{{\partial {\tilde\varepsilon _{xy}}}} = a\dfrac{{\partial {I_1}}}{{\partial {\tilde\varepsilon _{xy}}}} + \dfrac{{0.5b}}{{\sqrt {cI_1^2 + dJ_2^{'}} }}\left( {2c{I_1}\dfrac{{\partial {I_1}}}{{\partial {\tilde\varepsilon _{xy}}}} + d\dfrac{{\partial J_2^{'}}}{{\partial {\tilde\varepsilon _{xy}}}}} \right)
    \end{equation}
    with $\dfrac{{\partial {I_1}}}{{\partial {\tilde\varepsilon _{xx}}}} = 1 + \dfrac{{\partial {\tilde\varepsilon _{zz}}}}{{\partial {\tilde\varepsilon _{xx}}}}$, $\dfrac{{\partial {I_1}}}{{\partial {\tilde\varepsilon _{yy}}}} = 1 + \dfrac{{\partial {\tilde\varepsilon _{zz}}}}{{\partial {\tilde\varepsilon _{yy}}}}$, $\dfrac{{\partial {I_1}}}{{\partial {\tilde\varepsilon _{xy}}}} = 0$,\\
    $\dfrac{{\partial J_2^{'}}}{{\partial {\tilde\varepsilon _{xx}}}} = \dfrac{1}{3}\left( {2{\tilde\varepsilon _{xx}} + 2{\tilde\varepsilon _{zz}}\dfrac{{\partial {\tilde\varepsilon _{zz}}}}{{\partial {\tilde\varepsilon _{xx}}}} - {\tilde\varepsilon _{yy}} - {\tilde\varepsilon _{zz}} - \dfrac{{\partial {\tilde\varepsilon _{zz}}}}{{\partial {\tilde\varepsilon _{xx}}}}\left( {{\tilde\varepsilon _{xx}} + {\tilde\varepsilon _{yy}}} \right)} \right)$\\
    $\dfrac{{\partial J_2^{'}}}{{\partial {\tilde\varepsilon _{yy}}}} = \dfrac{1}{3}\left( {2{\tilde\varepsilon _{yy}} + 2{\tilde\varepsilon _{zz}}\dfrac{{\partial {\tilde\varepsilon _{zz}}}}{{\partial {\tilde\varepsilon _{yy}}}} - {\tilde\varepsilon _{xx}} - {\tilde\varepsilon _{zz}} - \dfrac{{\partial {\tilde\varepsilon _{zz}}}}{{\partial {\tilde\varepsilon _{yy}}}}\left( {{\tilde\varepsilon _{xx}} + {\tilde\varepsilon _{yy}}} \right)} \right)$\\
    $\dfrac{{\partial J_2^{'}}}{{\partial {\tilde\varepsilon _{xy}}}} = 2{\tilde\varepsilon _{xy}}$
\end{itemize}

\bibliographystyle{elsarticle-num}
\bibliography{HaiBib}
%
%
%

\end{document}